\documentclass[lettersize,journal,onecolumn,pdflatex]{IEEEtran}

\usepackage{graphicx}
\usepackage{subfig}
\usepackage{pict2e}
\usepackage{cite}
\usepackage{fullpage}
\usepackage{microtype}
\usepackage{colonequals}

\usepackage{amsopn}
\usepackage{amssymb}
\usepackage{theorem}
\usepackage{color}

\newcommand{\A}{{\cal A}}

\newcommand{\qed}{\hfill\hbox{\rule{6pt}{6pt}}}
\newtheorem{theorem}{Theorem}

\newtheorem{cor}[theorem]{Corollary}
\newtheorem{prop}[theorem]{Proposition}

\theorembodyfont{\rmfamily}
\theorembodyfont{\rmfamily}\newtheorem{example}[theorem]{Example}
\DeclareMathOperator{\pow}{Pow}

\renewcommand{\baselinestretch}{2}

\renewcommand\refname{References}
\renewcommand\figurename{Figure}

\renewcommand{\abstractname}{Abstract}

\usepackage[utf8]{inputenc} 
\usepackage[T1]{fontenc}
\usepackage{url}
\usepackage{ifthen}
\usepackage{cite}
\usepackage[cmex10]{amsmath} 

\begin{document}
\title{The explicit formulae for the distributions of nonoverlapping  words and its applications to statistical tests for
pseudo random numbers}

 \author{%
   \IEEEauthorblockN{Hayato Takahashi}
   \IEEEauthorblockA{Random Data Lab. Inc.\\
                     1210062 Tokyo Japan\\
                     Email: hayato.takahashi@ieee.org}
   \thanks{Part of the paper has been presented at Probability Symposium 2018 Kyoto Univ., SITA2018,
Ergodic Theory and Related Fields 2018, MSJ Okayama 2018, Bernoulli-IMS One World 2020, IBMP2020, IEICE IT2021-123, and arXiv:2105.05172.}
 }

\maketitle

\begin{abstract}
The distributions of the number of occurrences of words (the distributions of words for short) play key roles 
in information theory, statistics, probability theory, ergodic theory, computer science, and DNA analysis.
 Bassino et al.~2010 and Regnier et al.~1998 showed 
generating functions of the distributions of words for all sample sizes.
Robin et al.~1999 presented  
generating functions of the distributions for the return time of words and demonstrated a recurrence formula for these distributions.
These generating functions are rational functions; except for simple cases, it is difficult to expand them into power series. 
In this paper, we study finite-dimensional  generating functions of the distributions of nonoverlapping words for each fixed sample size 
and demonstrate the explicit formulae for the distributions of words for the Bernoulli models.
Our results are generalized to nonoverlapping partial words.
We study statistical tests that depend on the number of occurrences of words and
the number of block-wise occurrences of words, respectively. We demonstrate that the power of the test that depends on the number of occurrences of words is significantly large compared to the other one. 
Finally, we apply our results to statistical tests for pseudo random numbers.
\end{abstract}

\section{Introduction}
We study the distributions of the number of occurrences of  words (the distribution of words for short) of the finite alphabet,
which play key roles in statistical science, probability theory,  information theory, ergodic theory, computer science, and DNA analysis, 
see \cite{{JaquetandSzpankowski},{waterman},{bertheRigo},{lothire},{bookRobin},{Robin99},{li},{LZ77},{Ziv78},{shields96},{nielsen}}.

Classical studies on the distributions of words \cite{{Guibas81},{RegnierSpankowski98},{JaquetandSzpankowski},{Goulden83},{Bassinoetal2011},{feller1},{Robin99},{waterman}} concern
the generating functions of the distributions of words for all sample sizes and
present them as rational functions.
The asymptotic formulae for the distributions of
words are obtained from rational generating functions \cite{{JaquetandSzpankowski},{Flajolet2009}}.
Robin et al.~1999 \cite{Robin99}  studied 
rational generating functions of the distributions for the return time of words and demonstrated a recurrence formula for these distributions.

For example, let \(F(z)\colonequals\sum_n f(n)z^n\), where \(f(n)\) is the number of binary strings of length \(n\) that do not contain
the word \(01\). 
The generating function \(F\) is given by (\ref{eqGuibas}) for \(|z|<1\) (a special case of the theorem in Guibas and Odlyzko \cite{Guibas81}, see also pp.61 \cite{Flajolet2009}).
We have
\begin{align}
F(z) &=\frac{1}{1-2z+z^2}\label{eqGuibas}\\
&=\frac{1}{(1-z)^2}=(\sum_n z^n)^2=\sum_n (n+1)z^n\nonumber
\end{align}
and \(f(n)=n+1\).
For example, \(f(3)=4\) and the four words, 000, 100, 110, and 111 do not contain the word 01 among the strings of length 3.

However, it is difficult to expand rational functions into power series except for simple cases.
 For example, to obtain the coefficient of the expansion of rational functions by the partial fraction expansion, 
we need to obtain the all roots of the denominator and it is impractical to do so except for simple cases, see pp.~275 Feller \cite{feller1}.

We study the finite-dimensional  generating functions for the probabilities of the number  of nonoverlapping words 
for each fixed sample size
and present   the  explicit  formulae for the exact distributions of nonoverlapping words  
(Theorem~\ref{th-main2}). We give all orders of moments of the distributions nonoverlapping words (Theorem~\ref{th-moments}).
In Section~\ref{secPartialWords}, our results are generalized to nonoverlapping partial words.
In Section~\ref{secPowfunc}, we study statistical tests that depend on the number of occurrences of words and
the number of block-wise occurrences of words, respectively. We show that the power of the test that depends on the number of occurrences of words is significantly large compared to the other one. 
In Section~\ref{sec-pseudo},  we apply our results to statistical tests for pseudo random numbers.

\section{Distributions of nonoverlapping words}\label{secMain}
Let \(X^n\colonequals X_1\ldots X_n\) be random variables that take value in finite alphabet \(\A\) and
 \(N(w_1,\ldots,w_l; X^n)\)  the number of the appearances of the words \(w_1,\ldots,w_l\) in an arbitrary position of \(X^n\).
For example \(N(10,11;1011101)=(2,2)\).
Let \(|x|\) be the length of \(x\). 
A word \(x\) is called overlapping if there is a word \(z\) such that \(x\) appears at least 2 times in \(z\) and \(|z|<2|x|\) otherwise
\(x\) is called nonoverlapping. 
A pair of words \(x,y\) is called overlapping if there is a word \(z\) such that \(x\) and \(y\) appear in \(z\) and \(|z| < |x|+|y|\)
otherwise, the pair is called nonoverlapping.
A word \(x\) is overlapping if and only if \((x,x)\) is overlapping. 
A finite set of words \(S\) is called nonoverlapping if  every pair \((x,y)\) for \(x,y\in S\) are nonoverlapping,
otherwise, \(S\) is called overlapping. 
For example, sets of words, \(\{11\},\ \{10,01\},\ \text{and }\{00,11\}\) are overlapping, and \(\{10\}\text{ and } \{00111,00101\}\) are nonoverlapping.
\begin{theorem}\label{th-main2}
Let \(X_1X_2\cdots X_n\)  be an i.i.d.~process with finite alphabet.
Let \(w_1,\ldots, w_l\) be the set of nonoverlapping words.
Let  \(m_i\colonequals|w_i|\) and \(P(w_i)\) be the probability of \(w_i\) for \(i=1,\ldots, l\). 
Let 
\begin{align}
&A(k_1,\ldots,k_l)=\dbinom{n-\sum_i m_i k_i + \sum_i k_i}{k_1,\ldots, k_l}\prod_{i=1}^l P^{k_i}(w_i),\nonumber\\
&B(k_1,\ldots,k_l)=P(\sum_{i=1}^n I_{X_i^{i+m_i-1}=w_j}=k_j,\ j=1,\ldots,l),\label{eq-A}\\
&F_A(z_1,\ldots,z_l)=\sum_{k_1,\ldots,k_l}A(k_1,\ldots,k_l)z^{k_1}\cdots z^{k_l},\text{ and}\nonumber\\
&F_B(z_1,\ldots,z_l)=\sum_{k_1,\ldots,k_l}B(k_1,\ldots,k_l)z^{k_1}\cdots z^{k_l}.\nonumber
\end{align}
 Then
\[
F_A(z_1,z_2,\ldots,z_l)=F_B(z_1+1, z_2+1,\ldots,z_l+1),
\]
and
\begin{align}\label{eqmain}
& P(N(w_1,\ldots,w_l;X^n)=(s_1,\ldots,s_l))\nonumber\\
& =\sum_{\substack{k_1,\ldots,k_l\colon\\ \ s_1\leq k_1,\ldots,s_l\leq k_l\\ \sum_i m_i k_i\leq n}} (-1)^{\sum_i k_i-s_i}
\nonumber\\
& \dbinom{n-\sum_i m_i k_i + \sum_i k_i}{s_1,\ldots, s_l,k_1-s_1,\ldots k_l-s_l}\prod_{i=1}^l P^{k_i}(w_i) .
\end{align}
\end{theorem}
Proof)
For simplicity, we prove the theorem for \(l=1\).
The proof of the general case is similar. 
Let \(m=|w|\).
Since  \(w\) is nonoverlapping,
the number of possible allocations such that \(k\) times the appearance of \(w\)  in the string of length \(n\)  is
\[\dbinom{n-mk+k}{k}.\]
This is because if we replace  each \(w\)  with additional extra  symbol \(\alpha\) in the string of length \(n\) then the problem  reduces to 
choosing  \(k\)  \(\alpha\)'s   among the string of length  \(n-mk+k\).
Let
\begin{equation}\label{basicFuncA}
A(k)\colonequals\dbinom{n-mk+k}{k}P^{k}(w).
\end{equation}
The function \(A\) is not the probability of   \(k\) \(w\)'s occurrences in the string, since 
we allow any letters in the remaining place except for the appearance of  \(w\).
The function \(A\) may count the event that 
 \(w\) appears more than \(k\) times.
Let \(B(t)\) be the probability that \(w\) appears \(k\) times.
We have the following identity,
\[
A(k) =\sum_{k\leq t} B(t) \dbinom{t}{k}.
\]
Let \(F_A(z)\colonequals\sum_{k}A(k) z^{k}\) and 
\(F_B(z)\colonequals\sum_{k}B(k) z^{k}\).
Then
\begin{align*}
F_A(z) & =\sum_k z^k \sum_{k\leq t} B(t) \dbinom{t}{k}\\
& = \sum_{t} B(t) \sum_{k\leq t} \dbinom{t}{k} z^k\\
&= \sum_{t} B(t) (z+1)^{t} \\
& = F_B(z+1).
\end{align*}
We have
\begin{align*}
&F_B(z) =F_A(z-1)\\
&=\sum_{k\colon mk\leq n} \dbinom{n-mk+k}{k}(z-1)^kP^k(w)\\
&=\sum_{\substack{k,t\colon mk\leq n\\ t\leq k}}  \dbinom{n-mk+k}{k}\dbinom{k}{t}z^t(-1)^{k-t} P^k(w)\\
&=\sum_t z^t\sum_{\substack{k\colon mk\leq n\\ t\leq k}}  (-1)^{k-t} \dbinom{n-mk+k}{t,k-t} P^k(w),
\end{align*}
and  (\ref{eqmain}) .
\qed

Regnier et. al \cite{RegnierSpankowski98} showed expectation, variance, and central limit theorems (CLTs) for the occurrences of words.
We give all orders of moments for nonoverlapping words.
Let \(A_{t,s}\colonequals\sum_r \dbinom{s}{r}r^t(-1)^{s-r}\) for all \(t,s=1,2,\ldots\).
Then \(A_{t,s}\) is the number of surjective functions from \(\{1,2,\ldots,t\}\to\{1,2,\ldots,s\}\) for all \(t,s\), see pp.100 Problem 1  \cite{Riordan}.
\begin{theorem}\label{th-moments}
Let \(w\) be a nonoverlapping word.
\[E(N^t(w;X^n))=\sum_{s=1}^{\min \{T,t\}} A_{t,s}\dbinom{n-s|w|+s}{s}P^s(w)\]
for all \(t=1,2,\ldots\), where \(T=\max \{t \mid  t|w|\leq n\}\).
\end{theorem}
Proof)
Let
\(Y_i=I_{X_i^{i+|w|-1}=w}\).
We say that \(\{i,i+1,\ldots,i+|w|-1\}\) is the support of \(Y_i\).
Random variables \(Y_{n(1)},\ldots,Y_{n(s)}\)  are called disjoint if their support are disjoint, where
\(1\leq n(1),\ldots, n(s)\leq i+|w|-1\).
Since \(w\) is nonoverlapping, we have 
\begin{equation}\label{expect}
E(Y_iY_j)=
\begin{cases}
P(w) &\text{ if }i=j,\\
P^2(w)& \text{ if }Y_i\text{ and }Y_j\text{ are disjoint,}\\
0 &\text{ else.}
\end{cases}
\end{equation}

Let \(Y_{j,i}=Y_i\) for all \(1\leq j\leq t\).
Then  
\begin{align}
E(N^t(w;X^n))&=E((\sum_i Y_i)^t)=E(\prod_{j=1}^t\sum_i Y_{j,i})\nonumber\\
&=E(\sum_{n(1),\ldots,n(t)}\prod_{j=1}^t Y_{j,n(j)}).\label{subeq}
\end{align}
By (\ref{expect}), \(E( \prod_{j=1}^t Y_{j,n(j)})=P^s(w)\) if and only if  there is a disjoint set \(Y_{l(1)},\ldots,Y_{l(s)}\)
such that \(\{Y_{1,n(1)},\ldots,Y_{t,n(t)}\}=\{Y_{l(1)},\ldots,Y_{l(s)}\}\).

The number of possible combination of \(s\) disjoint \(\{Y_{l(1)},\ldots,Y_{l(s)}\}\) is \(\dbinom{n-s|w|+s}{s}\).
If \(n<s|w|\) then there is no \(s\) disjoint \(Y_i\).
For each disjoint \(\{Y_{l(1)},\ldots,Y_{l(s)}\}\),
the number of possible combination of \(n_1,\ldots, n_t\) such that \\
\(\{Y_{1,n(1)},\ldots,Y_{t,n(t)}\}=\{Y_{l(1)},\ldots,Y_{l(s)}\}\) is
\(A_{t,s}\). 
By (\ref{subeq}), we have the theorem.
\qed
\begin{example} Consider the three bits strings.
Then 000, 100, 110, and 111 have no \(01\) and the others have one  01.
On the other hand,
let \(P\) be a fair coin-flipping, \(n=3\), and \(w=01\) in Theorem~1.
Then
\[P(N(01;X^3)=s)=\sum_{s\leq k, 2k\leq 3} (-1)^{k-s}\dbinom{3-k}{s,k-s}2^{-2k}.\]
We have \(P(N=0)=1/2\) and \(P(N=1)=1/2\).
\end{example}

\section{Distributuions of nonoverlapping partial words}\label{secPartialWords}
We introduce the symbol  \(?\) to represent arbitrary symbols. 
Let \(\A\) be a finite alphabet and \(\A^*\) the set of words consisting of alphabet \(\A\).
A word consists of extended alphabet  \(\A\cup\{?\}\) is called {\it partial word} \cite{berstelBoasson}.
We say the word \(w'\in\A^*\) {\it realization} of the partial word \(w\) if \(w'\) coincides with \(w\) 
except for the symbol \(?\).
A partial word is called nonoverlapping if the set of the realization is nonoverlapping.
For example, \(001? 1\) is a nonoverlapping partial word with \(\A=\{0,1\}\) and its realizations are  00101 and 00111.

The probability of partial word \(w\) is defined by
\[P(w)\colonequals\sum_{w'\colon\text{realization of }w}P(w').\]
For example, \(P(1?1)\colonequals P(101)+P(111)\) if  \(\A=\{0,1\}\).
\begin{cor}
Theorem~\ref{th-main2} and Theorem~\ref{th-moments} holds for nonoverlapping partial words.
\end{cor}
Proof) The proof is the same as Theorem~\ref{th-main2} and Theorem~\ref{th-moments}.\qed

We can find many nonoverlapping partial words.
For example, \(0^{m+1}(1?^m)^n 1\) are nonoverlapping for all \(n,m\).
Here \(w^m\) is the \(m\) times concatenation of the word \(w\).
For example, \(0^3(1?^2 )^21=0001?? 1?? 1\).
We can construct large-size partial words that have large probabilities.
\begin{prop}\label{partialPropB}
Let  \(\A\colonequals\{0,1\}\), \(P\)  the fair-coin flipping, and 
\begin{equation}
 w(m)\colonequals 0^{m}(1?^{m-1})^{m-1} 1.
\end{equation}
Then \(w(m)\) is nonoverlapping for all \(m\),
\(|w(m)|=m^2+1\), and \(P(w(m))=2^{-2m}\).
\end{prop}
Proof)  Since  \(0^m\) is not overlapping for
all realization of  \(1?^{m-1}1\), \(w(m)\) is nonoverlapping for all \(m\).
The latter part is obvious. \qed

\section{Power function}\label{secPowfunc}
In this section, we study tests of statistical hypothesis:
Null hypothesis is \(P(w)=\theta^*\) vs alternative hypothesis is \(P(w)<\theta^*\),
where \(w\) is a nonoverlapping word and
\(0<\theta^*\leq 1\) is a fixed constant.

Let \(\A\colonequals \{0,1\}\),
 \(N_w\colonequals N(w; X^n)\), and
\[N'_w\colonequals \sum_{i=1}^{\lfloor n/|w|\rfloor} I_{X_{i*|w|}^{(i+1)*|w|-1}=w}.\]
We compare the power of tests that depend on \(N_w\) and \(N'_w\).

Let
\begin{align*}
&E(N_w)=(n-|w|+1)P(w)\text{ and} \\
&V(N_w)=E(N_w)+(n-2|w|+2)(n-2|w|+1)P^2(w).
\end{align*}
Then we have CLTs for the occurrences of words \cite{{Billingsley},{RegnierSpankowski98}}.
For all \(x\)
\[\lim_{n\to\infty}P(\frac{N_w-E(N_w)}{\sqrt{V(N_w)}}<x)= \frac{1}{\sqrt{2\pi}}\int^x_{-\infty} e^{-\frac{1}{2}x^2} dx.\]

By CLT, we construct tests that depend on \(N_w\) and \(N'_w\), respectively.
First, let \(E_\theta\colonequals E(N_w)\) and  \(V_\theta\colonequals V(N_w)\) if \(\theta=P(w)\).
Reject null hypothesis if and only if \(N_w<E_{\theta^*} -5\sqrt{V_{\theta^*}}\).

Next, 
let \(E'_\theta = \lfloor n/|w|\rfloor \theta\) and \(V'_\theta = \lfloor n/|w|\rfloor \theta (1-\theta)\) if 
\(\theta=P(w)\).
Reject null hypothesis if and only if \(N'_w<E'_{\theta^*} -5\sqrt{V'_{\theta^*}}\).
  
Let
\begin{align*}
&\pow(\theta)\colonequals  P_\theta(N_w<E_{\theta^*} -5\sqrt{V_{\theta^*}})\text{ and}\\
&\pow^{\prime}(\theta)\colonequals  P_\theta(N'_w<E'_{\theta^*} -5\sqrt{V'_{\theta^*}})\text{ for }\theta\leq \theta^*.
\end{align*}
The functions \(\pow\) and \(\pow^\prime\) are called power function.

\begin{figure}[h]
\scalebox{0.45}
{\includegraphics[0in,0in][4in,4in]{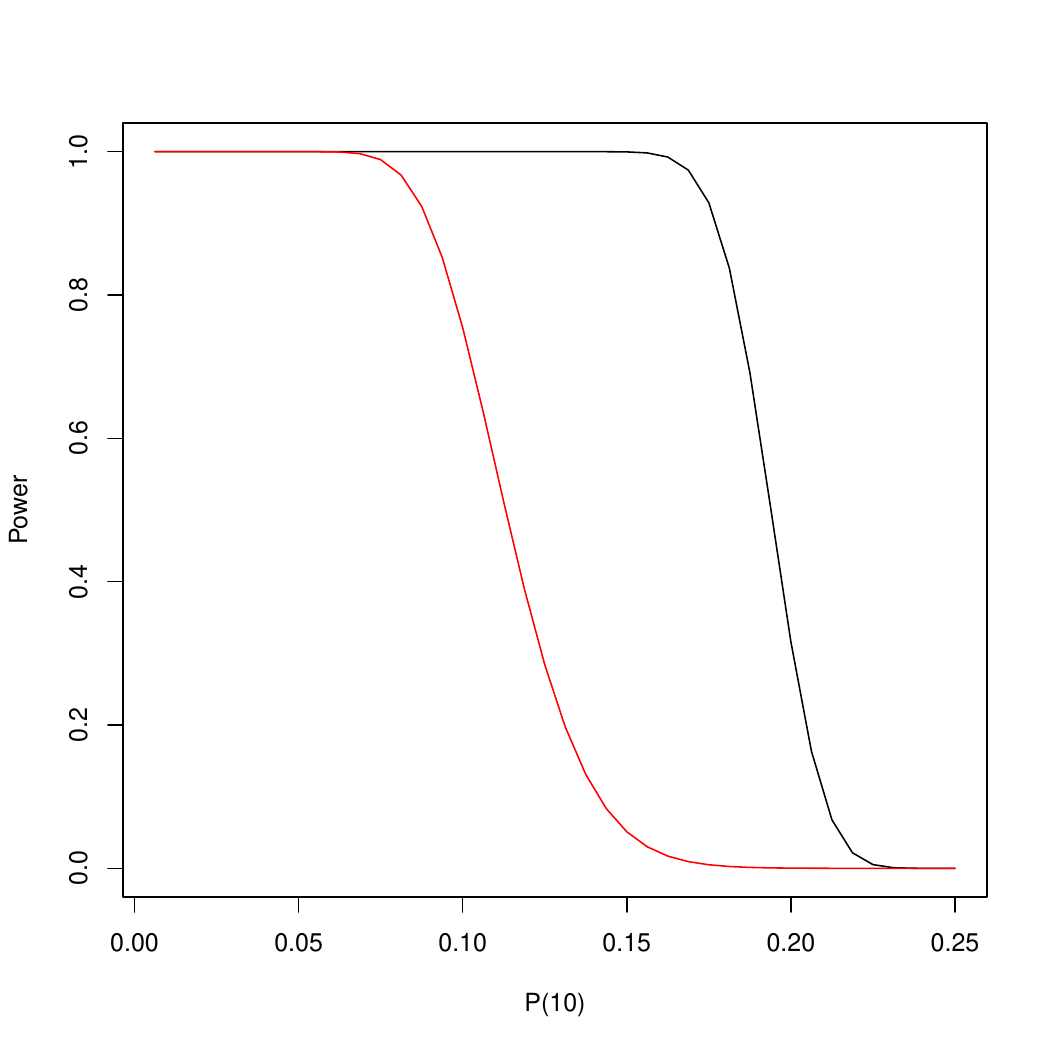}}
\caption{The graph of power functions for null hypothesis \(P(10)=0.25\) with \(n=500\).
The black line is the graph of \(\pow(P(10))\) and the red line is that of \(\pow^\prime(P(10))\).
}\label{fig-A}
\end{figure}

Let \(w=10\), \(\theta^*=0.25\), and \(n=500\).
The following table shows \(\pow(P(10))\) and  \(\pow^\prime(P(10))\) at \(P(10)=0.2\) and \(0.175\), respectively.
Figure~\ref{fig-A}  shows the graph of \(\pow(P(10))\) and  \(\pow^\prime(P(10))\).
We see that the power of the test that depends on \(N_w\) is significantly large compared to that of \(N'_w\).
\begin{center}
  \begin{tabular}{@{} cccc@{}}
    \hline
\(P(10)\) &0.2 &0.175 \\ 
    \hline
\(\pow(P(10))\) &0.316007 &0.928624 \\ 
\(\pow^\prime(P(10))\) &0.000295 &0.004982\\ 
    \hline
  \end{tabular}
\end{center}
\vspace{3cm}
\section{Experiments on statistical tests for pseudo random numbers}\label{sec-pseudo}
In  \cite{NIST2010}, a battery of statistical tests for pseudo random number generators is proposed, and 
the chi-square test is recommended to test the pseudo random numbers with
  \(N_w\)  for a nonoverlapping word \(w\).

\begin{figure}[ht]
\scalebox{0.44}{
\includegraphics[0in,0in][4in,4in]{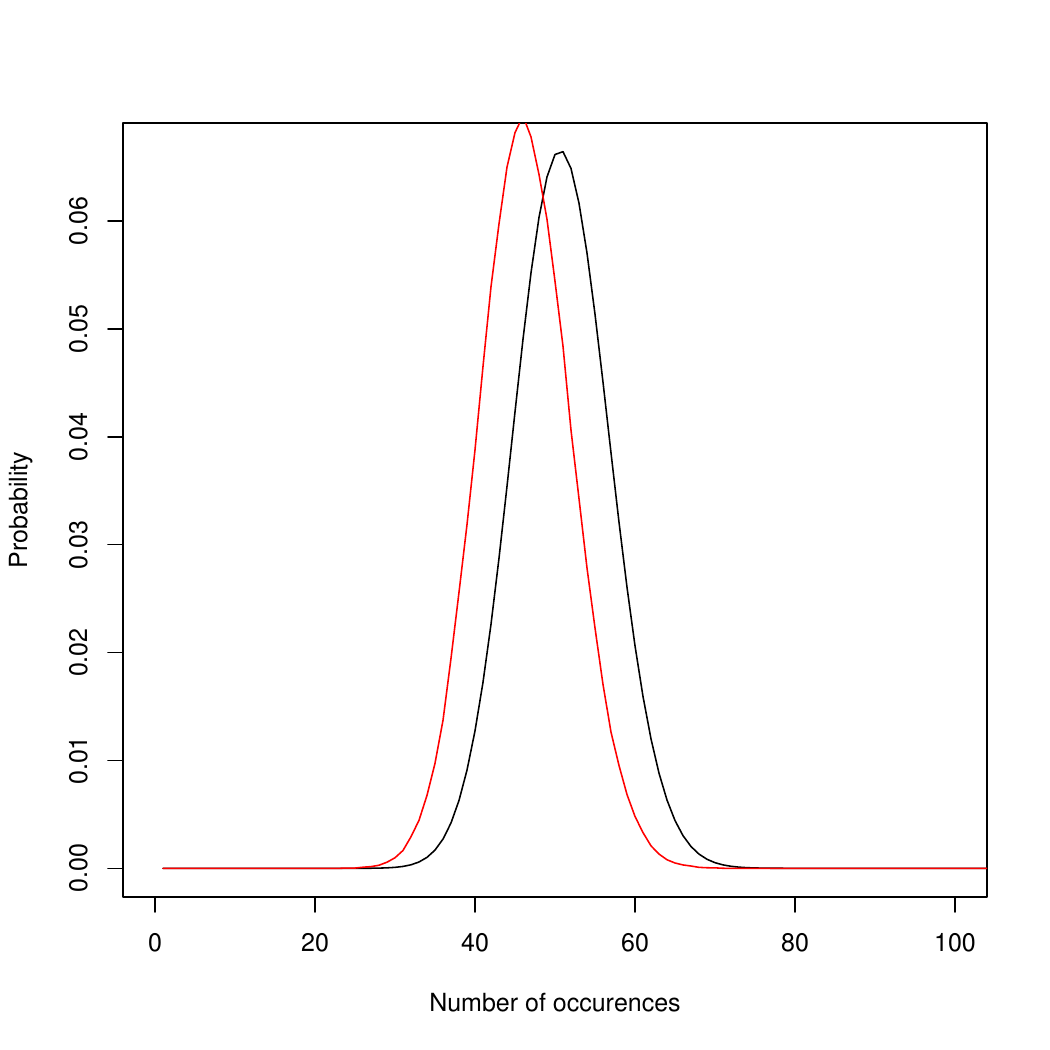}}
\caption{Comparison of distributions: 
The black line is the distribution of the number of occurrences of the word  \(11110\) in sample size  \(1600\).
The Red line is the corresponding empirical distribution  generated by
 the Monte Carlo simulation  with
 200000 BSD RNG random samplings.}\label{fig-dist}
\end{figure}

In this section, we  apply the Kolmogorov-Smirnov test to the empirical distributions of pseudo random numbers with 
 the true distributions of \(N_w\)  for a nonoverlapping word \(w\).
Let sample size \(n= 1600\) and \(l=1\) in  (\ref{eqmain}) and null hypothesis \(P\) be fair-coin flipping.
For each nonoverlapping words \(w=10\)  and \(w=11110\), 
we consider the  true distributions (\ref{eqmain}) of \(N_w\) and
 empirical distributions of \(N_w\) generated by the Monte Carlo method with BSD RNG.
 Figure~\ref{fig-dist} shows 
 the graph of these distributions for \(w=11110\).

\vspace*{3cm}

From the experiment for \(w=11110\), we have obtained
\[\sup_{0\leq x<\infty}| F_t(x)-F(x)|=0.302073,\]
where  \(F_t(x)\) is the empirical cumulative distribution generated by the Monte Carlo method with 
 \(t=200000\) BSD RNG  random samplings. 
\(F(x)\) is the true cumulative distributions of  (\ref{eqmain}).
The p-value of the Kolmogorov-Smirnov  test is almost zero,
\[P(\sup_{0\leq x<\infty}| F_t(x)-F(x)|\geq 0.302513 )\approx 0,\]
where \(P\) is the null hypothesis. 
The p-values of the Kolmogorov-Smirnov test for BSD RNG  with \(w=10,\ w=11110,\ t=200000\), and \(n=1600\) are  summarized in the following table.

\begin{center}
  \begin{tabular}{@{} cccc @{}}
    \hline
BSD RNG &w=10& w=11110 \\ 
    \hline
\(\sup_{0\leq x<\infty}| F_t(x)-F(x)|\) &0.012216 & 0.302073\\ 
p-value  & 0.000000 & 0.000000\\ 
    \hline
  \end{tabular}
\end{center}

The p-values of the Kolmogorov-Smirnov test for MT  RNG \cite{MT98} with \(w=10,\ w=11110,\ t=200000\), and \(n=1600\)  are  summarized in the following table.

\begin{center}
  \begin{tabular}{@{} cccc @{}}
    \hline
MT RNG &w=10& w=11110 \\ 
    \hline
\(\sup_{0\leq x<\infty}| F_t(x)-F(x)|\) & 0.001376& 0.001409\\ 
p-value  &0.843306  & 0.822066\\ 
    \hline
  \end{tabular}
\end{center}
\section*{Acknowledgments}
The author thanks for a helpful discussion with Prof. S.~Akiyama and Prof.~M.~Hirayama at Tsukuba University. This work was supported by the Research Institute for Mathematical Sciences, an International Joint Usage/Research Center located in Kyoto University, and Ergodic theory and related fields in Osaka University.

\end{document}